\documentstyle[aps,epsf,epsfig]{revtex}
\begin{document}
\input epsf
%\hskip7.5cm {UFES-DF-MMS.96/1}\\
%\baselineskip=25pt
%\draft
%\preprint
\title{Causality and Self-consistency in Classical  Electrodynamics }
\author{Manoelito M de Souza}
\address{Universidade Federal do Esp\'{\i}rito Santo - Departamento de
F\'{\i}sica\\29065.900 -Vit\'oria-ES-Brasil}
\date{\today}
\maketitle
\begin{abstract}
\noindent We present a pedagogical review of old inconsistencies of   Classical
Electrodynamics and of some new ideas that solve them. Problems with the
electron equation of motion and with the non-integrable singularity of its
self-field  energy tensor are well known. They are consequences, we show, of
neglecting terms that are null off the charge world-line but that give a non
null contribution on its world-line. The  electron self-field energy tensor is
integrable without the use of any kind of renormalization; there is  no
causality violation and no conflict with energy conservation in the electron
equation of motion, when its meaning is properly considered.
\end{abstract}
\begin{center}
\end{center}

\begin{section}
{INTRODUCTION}
\end{section}
\noindent Classical Electrodynamics of a point electron is based on the
Lienard-Wiechert solution;  its many old and unsolved problems
\cite{Rohrlich,Jackson,Parrot} make of it a non-consistent theory. One can
mention  the field singularity or the self-energy problem;  the non-integrable
singularities of its energy tensor;  the causality-violating behavior of
solutions of the Lorentz-Dirac equation \cite{Eliezer,Teitelboim,Rowe,Lozada};
etc. Here, we will discuss these problems. We will show that their solution is
connected to a more strict implementation of causality (extended causality)
which is explained in section II. In section III we review and discuss the
singularities and non-integrability of the electron self-field energy tensor.
Some helpful mathematical results are presented in section IV. They are useful
in the working out of some limiting processes. In section V the electron
equation of motion, which does not have the Schott term, is derived and its
physical meaning is discussed. Section VI is included like an appendix of
section V for showing an alternative way to the electron equation of motion
that illuminates its physical meaning. \\ The retarded Lienard-Wiechert solution
\begin{equation}
\label{LWS}
 A(x)=\frac{V}{\rho}{\Big|}_{{\tau}_{ret}},\;\;$for$\;\;\rho>0,
\end{equation}
is the retarded solution to the wave equation
\begin{equation}
\label{BoxA}
\Box A(x)=4\pi J(x)
\end{equation}
 and to
\begin{equation}
\label{dA}
\partial.A\equiv\frac{\partial A^{\mu}}{\partial x^{\mu}}=0,
\end{equation}
 where J, given by
\begin{equation}
\label{J}
J(x)=\int d\tau V\delta^{4}[x-z(\tau)],
\end{equation}
is the  current  for  a  point electron that describes a given trajectory
$z(\tau)$, parameterized by its proper-time $\tau;$ $V=\frac{dz}{d\tau}.$ The
electron charge and the speed of light are taken as 1.
\begin{equation}
\label{rho}
\rho:= -V_{\alpha}R^{\alpha}=-V.\eta.R=-V.R,
\end{equation}
 where $\eta$ is the  Minkowski metric tensor with signature +2,  and $R:=
x-z(\tau)$. $\rho$ is the invariant distance (in the charge rest frame) between
$z(\tau_{ret}),$ the position of the charge at the retarded time, and x, its
self-field event (See figure 1).
The constraints
 \begin{equation}
\label{lcone}
R^{2}=0,
\end{equation}
and
\begin{equation}
\label{rett}
R^{0}>0,
\end{equation}
 must be satisfied.
The constraint
$R^{2}=0$ requires that x and $z(\tau)$ belong to a same light-cone; it has two
solutions, $\tau_{ret}$ and $\tau_{adv}$, which are the points where J
intercepts the past and the future light-cone of x (see figure 1), and they
correspond, respectively to the advanced and the retarded solutions. The
retarded solution describes a signal emitted at $z(\tau_{ret})$ and that is
being observed at x, with $x^{0}>z^{0}(\tau_{ret})$, while the advanced
solution also observed at x, {\it will be} emitted in the future, at
$z(\tau_{adv})$, with $x^{0}>z^{0}(\tau_{adv})$. $R^{0}>0$ is a restriction to
the retarded solution (\ref{LWS}) as it excludes the causality violating
advanced solution, and justifies the restriction ${\big|}_{{\tau}_{ret}}$ in
(\ref{LWS}). But this is not the only available interpretation; we will show
below another one that does not have problems with causality violation and,
remarkably allows the description of particle  creation and annihilation still
in a classical physics context.\\

\begin{section}
{ Causality and spacetime geometry}
\end{section}

When working with variations or derivatives of A the constraint (\ref{lcone})
 must be considered in the neighbourhoods of x and of z: $x+dx$ and
$z(\tau_{ret}+d\tau)$ must also belong to a same light-cone.  A differentiation
 of (\ref{lcone}) ($R.dR=0\rightarrow R.(dx-Vd\tau)=0\rightarrow R.dx+\rho
d\tau=0$) generates the  constraint
\begin{equation}
\label{RdR}
d\tau+K.dx=0,
\end{equation}
 where K, defined for $\rho>0$, by
\begin{equation}
\label{K}
K:=\frac{R}{\rho},
\end{equation}
is a null 4-vector, $K^{2}=0$, and represents a light-cone generator, a tangent
to the light-cone.
The constraint (\ref{RdR}) defines a family of hyperplanes tangent to and
enveloped by the light-cone defined by $R^{2}=0$.   Together, these two
constraints require that x and $z(\tau_{ret})$ belong to a same straight line,
the x-lightcone generator tangent to $K^{\mu},$ or equivalently, orthogonal to
$K_{\mu}.$ See figures 1 and 2.\\
There is a geometric and physical interpretation of the two constraints
(\ref{lcone}) and (\ref{RdR}). $R^{2}=0$ assures that $A(x)$ is a signal that
propagates with the speed of light, on a light-cone; in field theory it
corresponds to the implementation of the so called {\it local causality}: only
points inside or on a same light-cone can be causally connected. It defines for
a physical object, at a point, {\it its physical spacetime}, that is the
regions of the space-time manifold that it can have access to.\\
But together, (\ref{lcone})  and (\ref{RdR}) produce a much more restrictive
constraint: a massless physical object cannot leave, by itself, its  light-cone
generator(labelled by K). Or, in other words, the part of a wavefront of $A(x)$
that moves along a light-cone generator must remain in this same generator.
This is in direct contradiction to  the Huyghens Principle that assumes that
the signal at a point of a wavefront is made of contributions from all points
of previous wavefronts; each point of a wavefront acts as a secondary source
emitting signal to all space directions. The Huyghens Principle is appropriate
for a description of light as a continuous wavy manifestation, but not for a
discrete one.\\
In contradistinction, the constraints (\ref{lcone}) and (\ref{RdR}), together,
imply that a point on a wavefront propagates, on its light-cone generator,
independently of all the other wavefront points. Each point of a wavefront,
therefore, can be treated as an entity by itself. It is so justified the naming
 of a CLASSICAL PHOTON to each point of an electromagnetic wavefront.
This corresponds to an EXTENDED CAUSALITY concept and it is readily extensible
to massive objects too \cite{hep-th/9505169}. It is appropriate for
descriptions of particle-like fields with discrete interactions, that is,
localized and propagating like a particle. Usually field theories are based on
a local-causality implementation, but it possible to build  a theory basing on
this extended causality. This is being discussed elsewhere \cite{CPMF}.\\
Armed with this extend-causality concept we can present another physical
interpretation of the two Lienard-Wiechert solutions. At the event x there are
two classical photons. One, that was emitted by the electron current $J$, at
$z(\tau_{ret})$ with $x^{0}>z^{0}(\tau_{ret})$, and is moving in the K
generator of the x-light-cone, $K^{\mu}:=(K^{0},{\vec K});$ . $J$ is its
source. The other one, moving on a ${\bar K}$-generator, ${\bar
K}^{\mu}:=(K^{0},-{\vec K}),$ will be absorbed by $J$ at $z(\tau_{adv}),$ with
$x^{0}<z^{0}(\tau_{adv}).$ $J$ is its sink. See figure 2. They are both
retarded solutions and correspond, respectively, to the creation and
destruction of a ``classical photon". Exactly this: creation and destruction of
particles in classical physics! This interpretation is only possible with these
concepts of extended causality and of classical photon.\\

\begin{section}
{ ENERGY TENSOR AND INTEGRABILITY}
\end{section}

When taking derivatives of $A(x)$ we must consider the restriction (\ref{RdR}),
or equivalently, $K_{\mu}=-\frac{\partial \tau_{ret}}{\partial x^{\mu}}.$ This
can turn, for the untrained, a trivial calculation into a mess. The best and
more fruitful approach, in our opinion, is to take $x$ and $\tau_{ret}$ as 5
independent parameters, and introduce a new derivative operator $\nabla,$
replacing the usual one:
\begin{equation}
\label{nabla}
\frac{\partial}{\partial
x^{\mu}}\Rightarrow\nabla_{\mu}:=\frac{\partial}{\partial
x^{\mu}}+\frac{\partial\tau}{\partial
x^{\mu}}\frac{\partial}{\partial\tau}=\frac{\partial}{\partial
x^{\mu}}-K_{\mu}\frac{\partial}{\partial\tau},
\end{equation}
or $\nabla_{\mu}:={\partial} _{\mu} -K_{\mu}\partial_{\tau},$ in a shorter
notation. The geometric meaning of $\nabla$ is quite clear; it is the
derivative allowed by the restriction (\ref{RdR}), that is, displacements on
the hyperplane $d\tau+K.dx=0$ only. The constraints (\ref{lcone}) and
(\ref{RdR}) together restricts $\nabla$
 to displacements along the K light-cone generator only.
Therefore, $\partial_{\mu}A(x)$, with the restriction ${\big|}_{{\tau}_{ret}}$
implicit is equivalent to $\nabla_{\mu}A(x)$ without any restriction.
\begin{equation}
\label{nablaa}
\partial_{\mu}A(x){\big|}_{{\tau}_{ret}}=\nabla_{\mu}A(x)
\end{equation}
 This corresponds to a geometrization of the extended causality concept.\\
Therefore we can write
\begin{equation}
\label{nablaA}
\nabla_{\mu}A^{\nu}=\nabla_{\mu}\frac{V^{\nu}}{\rho}=-\frac{K_{\mu}\hbox{\Large
%% FOLLOWING LINE CANNOT BE BROKEN BEFORE 80 CHAR
a}^{\nu}}{\rho}-\frac{V^{\nu}}{\rho^{2}}\nabla_{\mu}\rho=-K_{\mu}\frac{\hbox{\Large a}^{\nu}}{\rho}-\frac{V^{\nu}(K_{\mu}E-V_{\mu})}{\rho^{2}},
\end{equation}
with
\begin{equation}
\label{E}
E=1+\hbox{\Large a}.R=1+\rho\hbox{\Large a}_{K},
\end{equation}
as $\nabla_{\mu}V^{\nu}=-K_{\mu}\hbox{\Large a}^{\nu}$ and
\begin{equation}
\label{nablarho}
\nabla_{\mu}\rho=K_{\mu}E-V_{\mu},
\end{equation}
 where $\hbox{\Large a}_{K}:=\hbox{\Large a}.K.$ For notation simplicity we use
$[A,B]$ standing for  $\;[A_{\mu},B_{\nu}]:=A_{\mu}B_{\nu}-B_{\mu}A_{\nu}\;$
and (A,B)  for $(A_{\mu},B_{\nu}):=A_{\mu}B_{\nu}+A_{\nu}B_{\mu}$, and we are
omitting, from now on, the always implicit  restriction
${\big|}_{{\tau}_{ret}}.$ \\
We observe that the Lorentz gauge condition is automatically satisfied
\begin{equation}
\label{Lgc}
\nabla.A=-\frac{\rho\hbox{\Large a}_K+V.\nabla\rho}{\rho^{2}}=0,
\end{equation}
as $V.K=-1$, $V^{2}=-1$, and $V.\nabla\rho=1-E=-\rho\hbox{\Large a}_K$.\\
The Maxwell field ${F_{\mu\nu}}:=\nabla_{\nu}A_{\mu}-\nabla_{\mu}A_{\nu}$, is
found to be
\begin{equation}
\label{F}
{F}=\frac{1}{\rho^{2}}[K,W],
\end{equation}
with
\begin{equation}
\label{W}
W^{\mu}=\rho\hbox{\Large a}^{\mu}+EV^{\mu}.
\end{equation}
The electron self-field energy-momentum tensor,
$4\pi\Theta=F.F-\frac{\eta}{4}F^{2}$, is
\begin{equation}
\label{t}
%% FOLLOWING LINE CANNOT BE BROKEN BEFORE 80 CHAR
4\pi\rho^{4}\Theta^{\mu\nu}=[K^{\mu},W^{\alpha}][K_{\alpha},W^{\nu}]-\frac{{\eta}^{\mu\nu}}{4}[K^{\alpha},W^{\beta}][K_{\beta},W_{\alpha}],
\end{equation}
or in an expanded expression
\begin{equation}
\label{t'}
-4\pi\rho^{4}\Theta=(K,W)+KKW^{2}+WWK^{2}+\frac{\eta}{2}(1-K^{2}W^{2}),
\end{equation}
as $K.W=-1$.  We will use rather compact expressions like (\ref{t}) instead of
(\ref{t'}) also because they make easier the calculation of some limits that we
will have to do later. With $W^{2}=\rho^{2}\hbox{\Large
a}^{2}-E^{2}=\rho^{2}\hbox{\Large a}^{2}-(1+\rho\hbox{\Large a}_K)^{2},$
$\Theta$ may be written, according to its powers of $\rho,$ as
$\Theta=\Theta_{2}+\Theta_{3}+\Theta_{4}$, with
\begin{equation}
\label{t2}
4\pi\rho^{2}\Theta_{2}=[K,\hbox{\Large a}+V\hbox{\Large a}_K].[K,\hbox{\Large
a}+V\hbox{\Large a}_K]-\frac{\eta}{4}[K,\hbox{\Large a}+V\hbox{\Large
a}_{K}]^{2},
\end{equation}
or, $$-4\pi\rho^{2}\Theta_{2}=KK(\hbox{\Large a}^{2}-\hbox{\Large
a}_K^{2})+K^{2}(\hbox{\Large a}+V\hbox{\Large a}_K)(\hbox{\Large
a}+V\hbox{\Large a}_K)-\frac{\eta}{2}K^{2}(\hbox{\Large a}^{2}-\hbox{\Large
a}_K^{2}).$$
\begin{equation}
\label{t3}
4\pi\rho^{3}\Theta_{3}=[K,V].[K,\hbox{\Large a}+V\hbox{\Large
a}_K]+[K,\hbox{\Large a}+V\hbox{\Large
a}_K].[K,V]-\frac{\eta}{2}Tr[K,V].[K,\hbox{\Large a}],
\end{equation}
or $$4\pi\rho^{3}\Theta_{3}=-(K+VK^{2},\hbox{\Large a}+V\hbox{\Large
a}_K)+(2KK-\eta K^{2})\hbox{\Large a}_K.$$
\begin{equation}
\label{t4}
4\pi\rho^{4}\Theta_{4}=[K,V].[K,V]-\frac{\eta}{2}[K,V]^{2}.
\end{equation}
or $$4\pi\rho^{4}\Theta_{4}=KK-(K,V)-K^{2}VV-\frac{\eta}{2}(1+K^{2}).$$
If we neglect the $K^{2}$-terms in (\ref{t2}-\ref{t4}) we have:
\begin{equation}
\label{t2e}
4\pi\rho^{2}\Theta_{2}{\big|}_{K^{2}=0}=-KK{\biggl (}\hbox{\Large
a}^{2}-{\hbox{\Large a}_{K}}^{2}{\biggr )},
\end{equation}
\begin{equation}
\label{t3e}
4\pi\rho^{3}\Theta_{3}{\big|}_{K^{2}=0}=2KK\hbox{\Large a}_K -{\biggl (}
K,\hbox{\Large a}+V\hbox{\Large a}_K{\biggr )},
\end{equation}
\begin{equation}
\label{t4e}
4\pi\rho^{4}\Theta_{4}{\big|}_{K^{2}=0}=KK -(K,V) -\frac{\eta}{2},
\end{equation}
which are the usual expressions that one finds, for example in
\cite{Rohrlich,Jackson,Parrot,Teitelboim,Rowe,Lozada}.
Observe that
\begin{equation}
\label{C}
K.\Theta_{2}{\big|}_{K^{2}=0}=0,
\end{equation}
which is important in the identification of $\Theta_{2}$ with the radiated
\cite{Teitelboim} part of $\Theta$, and that
\begin{equation}
\label{C1} K.\Theta_{3}{\big|}_{K^{2}=0}=0.
\end{equation}
The presence of non-integrable singularities in the electron self-field energy
tensor is a major problem.
$\Theta_{2}{\big|}_{K^{2}=0}$, although singular at $\rho=0$, is nonetheless
integrable. By that it is meant that it produces a finite  flux through a
spacelike hypersurface $\sigma$ of normal $n$, that is,  $\int
d^{3}\sigma\Theta_{2}.n$ exists \cite{Rowe}, while
$\Theta_{3}{\big|}_{K^{2}=0}$ and $\Theta_{4}{\big|}_{K^{2}=0}$ are not
integrable; they generate, respectively, the problematic Schott term in the LDE
and a divergent term, the electron bound 4-momentum \cite{Teitelboim}, which
includes the so called electron self-energy.
Previous attempts, based on distribution theory, for  taming these
singularities have relied on modifications of the Maxwell
theory with  addition of extra terms to $\Theta{\bigg|}_{K^{2}=0}$ on the
electron world-line (see for example the reviews
\cite{Teitelboim,Rowe,Lozada}). They redefine $\Theta_{3}{\bigg|}_{K^{2}=0}$
and $\Theta_{4}{\bigg|}_{K^{2}=0}$ at the electron world-line in order to make
them integrable without changing them at $\rho>0,$  so to preserve the standard
results of Classical Electrodynamics. But this is always an ad hoc introduction
of something strange to the theory. Another unsatisfactory aspect of this
procedure is that it regularizes the above integral but leaves an unexplained
and unphysical discontinuity in the flux of 4-momentum, $\int
dx^{4}\Theta^{\mu\nu}\nabla_{\nu}\rho\;\delta(\rho-\varepsilon_{1}),$ through a
 cylindrical hypersurface $\rho=const$ enclosing  the charge world-line.
\noindent It is particularly interesting that, as we will show now, instead of
adding anything we should actually not drop out the null $K^2$-terms. Their
contribution (not null, in an appropriate limit) cancel the infinities. The
same problem happens in the derivations of the electron equation of motion from
these incomplete expressions of $\Theta.$  The Schott term in the Lorentz-Dirac
equation is a consequence; it does not
appear in the equation when the full expression of $\Theta$ is correctly used.

 By force of the constraints (\ref{lcone}) and (\ref{RdR}), as x and
$z(\tau_{ret})$ must remain on a same straight-line, the lightcone-generator K,
 the limit $\rho\rightarrow0$  necessarily implies also on $x^{\mu}\rightarrow
z(\tau_{ret})^{\mu}$ or $R^{\mu}\rightarrow 0.$\\
At $z(\tau_{ret}),\;\;\;K=\frac{R}{\rho}$ produces a $(\frac{0}{0})-$type of
indeterminacy, which can be evaluated at  neighboring points
$\tau=\tau_{ret}\pm d\tau$ by the L'Hospital rule and
$\frac{\partial}{\partial\tau}$ (see figure 3).
This application of the L'Hospital rule corresponds then to finding two
simultaneous limits: $\rho\rightarrow0$ and $\tau\rightarrow\tau_{ret}$.\\ As
\begin{equation}
\label{rodot}
\stackrel{.}{\rho}=-(1+\hbox{\Large a}.R)
\end{equation}
 and
\begin{equation}
\label{rdot}
\stackrel{.}{R}=-V,
\end{equation}
 then
\begin{equation}
\label{limitK}
\lim_{\rho\to0\atop \tau\to\tau_{ret}} K{\Big|}_{ R^{2}=0\atop R.dR=0} = V.
\end{equation}
This double limiting process is of course distinct of the single
$(\rho\rightarrow0)$-limit, which cannot avoid the singularity.
For notation simplicity we will keep using just $lim_{\rho\rightarrow0}$ but
always  with the implicit meaning as indicated in (\ref{limitK}). For example
by
\begin{equation}
\label{limitK2}
\lim_{\rho\to0}K^{2}=-1.
\end{equation}
we mean
\begin{equation}
\label{limitK2T}
\lim_{\rho\to0\atop \tau\to\tau_{ret}}K^{2}{\Big|}_{ R^{2}=0\atop R.dR=0}=-1.
\end{equation}
Classical Electrodynamics alone,  with its picture of a continuous emission of
radiation, does not give room for a comprehension of these limiting processes.
But we know that this classical continuous emission is just an approximate
description of an actually discrete quantum process. Figure 4 portrays a
classical picture (the electron and the photon trajectories) of such a
fundamental quantum process; it helps in the understanding of these two
limiting results. In the limit of $\rho\rightarrow0$  at $\tau=\tau_{ret}$
there are  3 distinct velocities: K, the photon 4-velocity, and $V_{1}$ and
$V_{2}$, the electron initial and final 4-velocities. This is the reason for
the indeterminacy at $\tau=\tau_{ret}$.
 At $\tau=\tau_{ret}+d\tau$ there is only $V_{2}$, and only $V_{1}$ at
$\tau=\tau_{ret}-d\tau.$  In other words, $\tau_{ret}$ is an isolated singular
point on the electron world-line; its neighboring $\tau_{ret}\pm d\tau$ are not
singular.  This is in flagrant contradiction to the Classical-Electrodynamics
assumption of a continuous emission process, because in this case, all points
in the electron world-line would be singular points, like $\tau_{ret}$.
It is remarkable that we can find vestiges of these traits of the quantum
nature of the radiation emission process in a classical  (Lienard-Wiechert)
solution. This is food for thinking on the real physical meaning of the
classical and the quantum fields.

\begin{section}
{ Some useful mathematical tools}
\end{section}

To find this double limit of something when $\rho\rightarrow0$ and
$\tau\rightarrow\tau_{ret}$ will be done so many times in this paper that it is
better to do it in a more systematic way.   We want to find
\begin{equation}
\label{LR}
\lim_{\rho\to0}\frac{N(R,\dots)}{\rho^{n}},
\end{equation}
where $N(R,\dots)$ is a homogeneous function of R,
$N(R,\dots){\big|}_{R=0}=0$. Then, we have to apply the L'Hospital rule
consecutively until the indeterminacy is resolved.  As
$\frac{\partial\rho}{\partial\tau}=-(1+\hbox{\Large a}.R)$, the denominator of
(\ref{LR}) at $R=0$ will be different of zero only after the
$n^{th}$-application of the L'Hospital rule, and then, its value will be
$(-1)^{n}n!$\\
If $p$ is the smallest integer such that $N(R,\dots)_{p}{\big|}_{R=0}\ne0,$
where $N(R)_{p}:=\frac{d^{p}}{d{\tau}^{p}}N(R,\dots)$, then
\begin{equation}
\label{NR}
\lim_{\rho\to0}\frac{N(R,\dots)}{\rho^{n}}=\cases{\infty,& if $p<n$\cr
              (-1)^{n}{\frac{N(0,\dots)_{p}}{n!}},& if $p=n$\cr
0,& if $p>n$\cr}
\end{equation}
\begin{itemize}
\item Example 1: $\cases{K=\frac{R}{\rho}.& $n=p=1 \Longrightarrow
\lim_{\rho\to0}K=V$\cr K^{2}=\frac{R.\eta.R}{\rho^{2}}.&$ n=p=2 \Longrightarrow
\lim_{\rho\to0}K^{2}=-1.$\cr}$
\item Example 2: $\frac{[K,\hbox{\Large a}]}{\rho}=\frac{[R,\hbox{\Large
a}]}{\rho^{2}}\;\;$ $\;\Longrightarrow \;$ $\;p=1<n=2\Longrightarrow
\lim_{\rho\to0}\frac{[K,\hbox{\Large a}]}{\rho}$ diverges
\item Example 3: $\frac{\hbox{\Large a}_K}{\rho}[K,V]=-\frac{\hbox{\Large
a}.R}{\rho^{3}}[R,V]\Longrightarrow p=4>n=3$ \qquad
$\lim_{\rho\to0}\frac{\hbox{\Large a}_K}{\rho}[K,V]=0$
\item Example 4: $\frac{[K,V]}{\rho^{2}}=\frac{[R,V]}{\rho^{2}}\;\;
\;\Longrightarrow
\;\;p=2<n=3\Longrightarrow\lim_{\rho\to0}\frac{[K,V]}{\rho^{2}}$ diverges
\end{itemize}
Finding these limits for more complex functions can be made easier with two
helpful expressions,
\begin{equation}
\label{NP}
N_{p}=\sum_{a=0}^{p}\pmatrix{p\cr a\cr}A_{p-a}.B_{a}
\end{equation}
and
\begin{equation}
\label{AgBC}
N_{p}=\sum_{a=0}^{p}\sum_{c=0}^{a}\pmatrix{p\cr a\cr}\pmatrix{a\cr
c\cr}A_{p-a}.B_{a-c}.C_{c}
\end{equation}
valid when $N(R)$ has, respectively, the forms $N_{0}=A_{0}.B_{0},$ or
$N_{0}=A_{0}.B_{0}.C_{0},$ where A, B and C represent possibly distinct
functions of R, and the subindices indicate the order of $\frac{d}{d\tau}$. For
example: $A_{0}=A$;  $A_{1}=\partial_{\tau} A$; $A_{2}=\partial_{\tau}^{2} A,$
and so on.
So, for using (\ref{NR}-\ref{AgBC}), we just have to find the
$\tau$-derivatives of A, B and C that produce the first non- null term at the
point limit of $R\rightarrow0.\;\;$\\
Consecutive derivatives of products of functions can become unwieldy. So it is
worthy to introduce the concept of ``$\tau$-order" of a function, meaning the
lowest order of the $\tau$-derivative of a function that produces a non-null
result at $R=0$. Let us represent the $\tau$-order" of f(x) by ${\cal
O}[f(x)].$ So, for example, from (\ref{rodot}) and (\ref{rdot}) we see that
\begin{equation}
\label{tor}
{\cal O}[R]=1,
\end{equation}
\begin{equation}
\label{torho}
{\cal O}[\rho]=1,
\end{equation}
As $\partial_{\tau}(\hbox{\Large a}.R)=-{\dot{\hbox{\Large a}}}.R$ and
$\partial^{2}_{\tau}(\hbox{\Large a}.R)=-{\ddot{\hbox{\Large
a}}}.R-{\dot{\hbox{\Large a}}}.V=-{\ddot{\hbox{\Large a}}}.R+\hbox{\Large
a}^{2}=\hbox{\Large a}^{2}+{\cal O}(R),$ then
\begin{equation}
\label{toar}
{\cal O}[\hbox{\Large a}.R]=2,
\end{equation}
For finding the $N_{p}$ of (\ref{NP}) and of (\ref{AgBC}) it is then necessary
to consider only the terms with the lowest $\tau$-order on each factor. Some
combinations of terms have derivatives that cancel parts of each other
resulting in a higher $\tau$-order term. For example,
$$\partial_{\tau}(R^{2}+\rho^{2})=+2\rho-2\rho E=-2\rho\hbox{\Large a}.R$$
$$\partial^{2}_{\tau}(R^{2}+\rho^{2})=2E\hbox{\Large
a}.R-2\rho{\dot{\hbox{\Large a}}}.R= 2(\hbox{\Large a}.R-\rho{\dot{\hbox{\Large
a}}}.R)+{\cal O}(R^{4}),$$
$$\partial^{3}_{\tau}(R^{2}+\rho^{2})=2({\dot{\hbox{\Large
a}}}.R+E{\dot{\hbox{\Large a}}}.R-\rho\hbox{\Large a}^{2})+{\cal
O}(R^{3})=4{\dot{\hbox{\Large a}}}.R-2\rho\hbox{\Large a}^{2}+{\cal
O}(R^{3}),$$
$$\partial^{4}_{\tau}(R^{2}+\rho^{2})=4\hbox{\Large a}^{2}+2\hbox{\Large
a}^{2}+{\cal O}(R^{2})=6\hbox{\Large a}^{2}+{\cal O}(R^{2}).$$
So, $${\cal O}[R^{2}+\rho^{2}]=4$$ although $${\cal O}[R^{2}]={\cal
O}[\rho^{2}]=2.$$
Observe that we only have to care with the lowest $\tau$-order terms as the
other ones, grouped in ${\cal O}(R)$, will not survive the limit
$R\rightarrow0$. Also, we do not care on writing the $\tau$-derivatives of
factors that will not reduce its $\tau$-order. For example in
$$\partial_{\tau}(RV+{\cal O}(R^{2}))=-VV+{\cal O}(R),$$  the term
$R\hbox{\Large a}$ was absorbed in ${\cal O}(R)$. In this way we avoid taking
unnecessary derivatives.
\begin{section}
{ The electron equation of motion}
\end{section}

\noindent The motion of a classical electron \cite{Rohrlich,Jackson,Parrot} is
described by the Lorentz-Dirac equation,
\begin{equation}
\label{LDE}
m\hbox{\Large a} = F_{ext}.V+\frac{2}{3}(\stackrel{.}{\hbox{\Large
a}}-\hbox{\Large a}^{2}V),
\end{equation}
where  m is the electron mass and $F_{ext}$ is an external electromagnetic
field.
The presence of the Schott term, $\frac{2}{3}e^{2}\!\stackrel{.}{a}$, is  the
cause  of  all of its pathological features, like microscopic non-causality,
runaway solutions, preacceleration, and other bizarre effects \cite{Eliezer}.
On the other hand, its  presence is apparently necessary for the
energy-momentum  conservation; without it it would be required a contradictory
null radiance for an accelerated charge, as  $\stackrel{.}{a}.V+a^{2}=0$. This
makes of the Lorentz-Dirac equation the greatest paradox of classical field
theory as it cannot simultaneously preserve both the causality and the energy
conservation \cite{Rohrlich,Jackson,Parrot}.\\
The Lorentz-Dirac equation can be obtained from energy-momentum conservation,
that leads to
\begin{eqnarray}
\label{eem}
m\hbox{\Large a}^{\mu}- F_{ext}^{\mu\nu}V_{\nu} & = &
mbox{}-\lim_{{\varepsilon_{1}\to0}\atop
{{\varepsilon_{2}\to\infty}\atop{\tau_{1}\to{\tau_{2}}}}}\int
%% FOLLOWING LINE CANNOT BE BROKEN BEFORE 80 CHAR
dx^{4}\nabla_{\nu}\Theta^{\mu\nu}\theta(\rho-\varepsilon_{1})\theta(\varepsilon_{2}-\rho)\theta(\tau_{2}-\tau)\theta(\tau-\tau_{1})= \nonumber \\
&=&\mbox{}
%% FOLLOWING LINE CANNOT BE BROKEN BEFORE 80 CHAR
-\lim_{{\varepsilon_{1}\to0}\atop{{\varepsilon_{2}\to\infty}\atop{\tau_{1}\to{\tau_{2}}}}}\int dx^{4}\Theta^{\mu\nu}{\Bigg(}\nabla_{\nu}\rho\;{\Big (}\theta(\rho-\varepsilon_{1})\delta(\varepsilon_{2}-\rho)
%% FOLLOWING LINE CANNOT BE BROKEN BEFORE 80 CHAR
-\delta(\rho-\varepsilon_{1})\theta(\varepsilon_{2}-\rho){\Big)}\theta(\tau_{2}-\tau)\theta(\tau-\tau_{1})+ \nonumber \\
& & \mbox{}
   + {\Big
%% FOLLOWING LINE CANNOT BE BROKEN BEFORE 80 CHAR
(}\theta(\tau_{2}-\tau)\delta(\tau-\tau_{1})-\delta(\tau_{2}-\tau)\theta(\tau-\tau_{1}){\Big )}{\Bigg)},
\end{eqnarray}
where $\;\tau_{2},\;\tau_{1},\;\varepsilon_{2},$ and $\varepsilon_{1}$ are
constants with $\;\tau_{2}>\tau_{1}$ and $\varepsilon_{2}>\varepsilon_{1}$.
$\theta(\rho-\varepsilon_{1})\theta(\varepsilon_{2}-\rho)$ defines the
spacetime region   between two coaxial  cylindrical Bhabha tubes surrounding
the electron world-line; for each fixed time they are reduced to two spherical
surfaces centred at the charge. $\theta(\tau_{2}-\tau)\theta(\tau-\tau_{1})$
defines the spacetime region   between two light-cones of vertices at
$\tau_{2}$ and $\tau_{1}$, respectively. They are necessary for using the
Gauss's theorem in the above intermediary step, as the product of these four
Heaviside functions define a closed hypersurface.  The terms in the second and
third lines of (\ref{eem}) are the flux rates of energy-momentum through the
respective hypersurfaces $\rho=\varepsilon_{1},$  $\rho=\varepsilon_{2},$
$\tau=\tau_{2}$ and $\tau=\tau_{1}$.\\
Taking the $\tau_{2}\rightarrow\tau_{1}$ limit we have
\begin{equation}
\label{feem}
m\hbox{\Large a}^{\mu}-
%% FOLLOWING LINE CANNOT BE BROKEN BEFORE 80 CHAR
F_{ext}^{\mu\nu}V_{\nu}=lim_{{\varepsilon_{1}\to0}\atop{\varepsilon_{2}\to\infty}}\int_{\tau_{2}} dx^{4}\Theta^{\mu\nu}\nabla_{\nu}\rho\;{\Big(}\theta(\rho-\varepsilon_{1})\delta(\varepsilon_{2}-\rho)-\delta(\rho-\varepsilon_{1})\theta(\varepsilon_{2}-\rho){\Big )}\delta(\tau-\tau_{2})
\end{equation}

Let us now apply (\ref{NR}-\ref{AgBC}) for finding
$\lim_{\rho\to0}\int_{\tau_{2}}
dx^{3}\Theta.\nabla\rho\delta(\rho-\varepsilon_{1}),$ which with the explicit
use of retarded coordinates (see, for example, p. 20 of \cite{Teitelboim}),
$x=z+\rho K,$ can be written as $\lim_{\rho\to0}\int_{\tau_{2}}
\rho^{2}d\rho\;d^{2}\Omega\;\Theta.\nabla\rho\;\delta(\rho-\varepsilon_{1}).$
In (\ref{t}), the definition of $\Theta$, the second term is the trace of the
first one and so we just have to consider this last one because the behaviour
of its trace under this limiting process can then easily be inferred. So, as
$K=\frac{R}{\rho},$ and $\nabla\rho=(KE-V)$ we have schematically, for the
first term of (\ref{t}) in $\rho^{2}\Theta.\nabla\rho$,
\begin{equation}
\label{10}
%% FOLLOWING LINE CANNOT BE BROKEN BEFORE 80 CHAR
\lim_{\rho\to0}\frac{N(R,\dots}{\rho^{n}}=\lim_{\rho\to0}\frac{\rho^{2}[K,W].[K,W].(KE-V)}{\rho^{4}}=\lim_{\rho\to0}\frac{[R,W].[R,W].(RE-V\rho)}{\rho^{5}}
\end{equation}
Then, comparing it with (\ref{LR}) and (\ref{AgBC}) we have
\begin{equation}
A_{0}=B_{0}=[R,W]=[R,\hbox{\Large a}\rho+VE]=[R,\hbox{\Large a}\rho+V]+{\cal
O}(R^{3})
\end{equation}
$$ A_{1}=B_{1}=[-V,\hbox{\Large a}\rho+V]+[R,-\hbox{\Large a}E+\hbox{\Large
a}]+{\cal O}(R^{2})=-[V,\hbox{\Large a}\rho]+{\cal O}(R^{2});$$
$$ A_{2}=B_{2}=-[\hbox{\Large a},V]+{\cal O}(R);$$
\begin{equation}
C_{0}=RE-V\rho=R-V\rho+{\cal O}(R^{3}),
\end{equation}
$$ C_{1}=-V-\hbox{\Large a}\rho+VE+{\cal O}(R^{2})=-\hbox{\Large a}\rho+{\cal
O}(R^{2}),$$
$$ C_{2}=\hbox{\Large a}+{\cal O}(R).$$
Therefore, for producing a possibly non null $N_{p}$, according to
(\ref{AgBC}),  $a, c$ and $p$  must be given by  $$c=2,$$
$$p-a=a-c=2\Longrightarrow p=6>n=5.$$
Or in a shorter way $${\cal O}[[R,W]]={\cal O}[RE-V\rho]=2,$$
and then, $$2{\cal O}[[R,W]]+{\cal O}[RE-V\rho]=6>n=5.$$
Then, we conclude from (\ref{NR}) that $N_{p}=0,$
\begin{equation}
\label{fpe}
\lim_{\rho\to0}\int_{\tau_{2}}
dx^{3}\Theta.\nabla\rho\delta(\rho-\varepsilon_{1})=0.
\end{equation}
The flux of energy and momentum rate of the electron self-field through the
$(\rho=\varepsilon_{1})$-hypersurface in (\ref{eem}) is null at
$\varepsilon_{1}=0.$  This is a new result, a consequence of (\ref{limitK}). In
the standard approach the contribution from this term produces the problematic
Schott term and a diverging expression, the electron bound-momentum which
requires  mass renormalization \cite{Teitelboim1}.\\ The RHS of (\ref{eem}) is
then reduced to
\begin{equation}
\label{fpr}
\lim_{{\varepsilon_{1}\to0}\atop{\varepsilon_{2}\to\infty}}\int_{\tau_{2}}
%% FOLLOWING LINE CANNOT BE BROKEN BEFORE 80 CHAR
dx^{3}\Theta^{\mu\nu}\nabla_{\nu}\rho\;\theta(\rho-\varepsilon_{1})\delta(\varepsilon_{2}-\rho)=\lim_{\varepsilon_{2}\to\infty}\int_{\tau_{2}}  \rho^{2}d\rho d^{2}\Omega\Theta^{\mu\nu}_{2}\nabla_{\nu}\rho\;\delta(\varepsilon_{2}-\rho),
\end{equation}
as, with (\ref{C1}), only $\Theta_{2}$ from $\Theta=\Theta_{2}
+\Theta_{3}+\Theta_{4}$ survives the passage $\varepsilon_{2}\to\infty$ in the
above integral. But from (\ref{nablarho}), (\ref{t2}) and (\ref{C}) we have
that
\begin{equation}
%% FOLLOWING LINE CANNOT BE BROKEN BEFORE 80 CHAR
\rho^{2}\Theta_{2}.\nabla\rho=-\rho^{2}V.\Theta_{2}=\frac{1}{4\pi}K(\hbox{\Large a}^{2}-\hbox{\Large a}_{K}^{2}).
\end{equation}
Then after the angular integration
\begin{equation}
\frac{1}{4\pi}\int d^{2}\Omega K(\hbox{\Large a}^{2}-\hbox{\Large
a}_{K}^{2})=\frac{2}{3}V^{\mu}\hbox{\Large a}^{2},
\end{equation}
 we have
\begin{equation}
\label{larmor}
\lim_{{\varepsilon_{1}\to0}\atop{\varepsilon_{2}\to\infty}}\int_{\tau_{2}}
%% FOLLOWING LINE CANNOT BE BROKEN BEFORE 80 CHAR
dx^{3}\Theta^{\mu\nu}\nabla_{\nu}\rho\;\delta(\varepsilon_{2}-\rho)=\frac{2}{3}V^{\mu}\hbox{\Large a}^{2}.
\end{equation}
The last passage is a well known (see, for example, page 111 of
\cite{Rohrlich}) text-book result; $-\frac{2}{3}V^{\mu}\hbox{\Large a}^{2}$ is
the Larmor term for the irradiated energy-momentum rate.\\
 With (\ref{fpe}) and (\ref{larmor}) in (\ref{feem}) we could  write the
electron equation of motion as
\begin{equation}
\label{fld}
m\hbox{\Large a}^{\mu}-F^{\mu\nu}_{ext}V_{\nu}=-\frac{2}{3}\hbox{\Large
a}^{2}V^{\mu},
\end{equation}
but it is well known that this could not be a correct equation because it is
not self-consistent: its LHS is orthogonal to V,
\begin{equation}
m\hbox{\Large a}.V=0\qquad\hbox{and}\qquad V.F_{ext}.V=0,
\end{equation}
 while its RHS is not,
\begin{equation}
-\frac{2}{3}\hbox{\Large a}^{2}V.V=\frac{2}{3}\hbox{\Large a}^{2}.
\end{equation}
 This seems to be paradoxical until one has a clearer idea of what is
happening. We must turn our attention to equation (\ref{feem}), where there is
a subtle and very important distinction between its LHS and its RHS. Its LHS is
entirely determined by the electron instantaneous position, $z(\tau),$ while
its RHS is determined by the sum of contributions from the electron self-field
at all points of a spherical surface.  In other words, the LHS is a description
of some electron attributes localized at a point (the electron position) while
the RHS  is a description of the sum of some electron-self-field attributes
over a spherical surface. This distinction is missing in equation (\ref{fld});
it was deleted by the integration process. The LHS of (\ref{feem}) multiplied
by V is null, we know, because the force that drives the electron with the
4-velocity V delivers a power ($m\hbox{\Large a}_{0}V^{0}$) that is equal to
the work per unit time realized by this force along the $\vec{V}$ direction
($m\vec{\hbox{\Large a}}.\vec{V}$) (this, we know, is the physical meaning of
$m\hbox{\Large a}.V=0$).  But this reasoning does not apply to the RHS of
(\ref{feem}) multiplied by V because the flux of  radiated energy is through a
spherical  surface $\rho=\varepsilon_{2},$ along K,  not along $V$ (except at
$\rho=0$, because of (\ref{limitK})); in order to make sense, as we are doing a
balance of the flux-rate of energy, we have to add this flux-rate from each
point  of the integration domain. Based on considerations of symmetry one can
anticipate that the final result must be null: to each point of a spherical
hypersurface $\rho=const.,\;\tau=\tau_{2},$ that gives a non-null contribution
there is another point giving an equal but with opposite sign contribution. The
RHS of (\ref{fld}) cannot be used for this point-to-point calculation as it
just represents a kind of average value. As a matter of fact, the equation
(\ref{fld}), in this sense, can be regarded as an effective equation that would
be better represented as
\begin{equation}
\label{finally}
m\hbox{\Large a}^{\mu}=F^{\mu\nu}_{ext}V_{\nu}-<\frac{2}{3}\hbox{\Large
a}^{2}V^{\mu}>,
\end{equation}
where the bracketed term represents the contribution from the electron
self-field:
 $$<\frac{2}{3}\hbox{\Large
a}^{2}V^{\mu}>=\lim_{{\varepsilon_{1}\to0}\atop{\varepsilon_{2}\to\infty}}\int
%% FOLLOWING LINE CANNOT BE BROKEN BEFORE 80 CHAR
dx^{3}\nabla_{\nu}\Theta^{\mu\nu}\theta(\rho-\varepsilon_{1})\theta(\varepsilon_{2}-\rho).\;$$ This is more than just a change of notation; it explicitly implies on a clear distinction between the V inside and the V outside the bracket in (\ref{finally}): $$<V>\neq V.$$
 Contributions from the electron self-field must always be calculated through
this point-by-point summation, like in the the RHS of (\ref{eem}) for the flux
of electromagnetic energy-momentum, through the walls of a Bhabha tube around
the charge world-line, in the limit of $\rho\rightarrow0$. In particular,
\begin{equation}
\label{dKt}
({m\hbox{\Large
%% FOLLOWING LINE CANNOT BE BROKEN BEFORE 80 CHAR
a}}-V.F_{ext}).V=-\lim_{{\varepsilon_{1}\to0}\atop{\varepsilon_{2}\to\infty}}\int_{\tau_{2}} d^{3}x  X_{\mu}\nabla_{\nu}\Theta^{\mu\nu}\theta(\rho-\varepsilon_{1})\theta(\varepsilon_{2}-\rho),
\end{equation}
where
\begin{equation}
\label{casesX}
X=\cases{K,&if $\rho>0$;\cr
V,&if $\rho=0$.\cr}
\end{equation}
X, in the RHS of (\ref{dKt}), gives the direction of the flux-rate of the
radiated energy; in the LHS the direction of the energy flux-rate is given by
V.
Observe that $X(\tau_{ret})$ is x-dependent and so it does not commute with
$\int d^{3}x$, that is, X inside and X outside the integral in the RHS of
(\ref{dKt}) give distinct results and, based on the above arguments, we are
saying that (\ref{dKt}) shows the correct way.
Its LHS is, of course, null. We show now that the RHS is also null, so that
there is no contradiction anymore.  We know that
\begin{equation}
\label{Vdt1}
%% FOLLOWING LINE CANNOT BE BROKEN BEFORE 80 CHAR
\nabla_{\nu}\Theta^{\mu\nu}=\frac{1}{4\pi}F^{\mu}_{\;\alpha}\nabla_{\nu}F^{\alpha\nu}=\frac{1}{4\pi}F^{\mu}_{\;\alpha}\Box A^{\alpha},
\end{equation}
and by direct calculation  we find that
\begin{equation}
\label{boxa}
\Box A^{\mu}=\frac{K^{2}}{\rho^{3}}{\Big(}3\rho E\hbox{\Large
a}^{\mu}+\rho^{2}{\dot{\hbox{\Large
a}}}^{\mu}+(3E^{2}+\rho^{2}{\dot{\hbox{\Large a}}}_{K})V^{\mu}{\Big )}.
\end{equation}
We see then that the integrand of the RHS of (\ref{dKt}) is null for $\rho>0$
as $K^{2}=0.$ For simplicity we could then just have used V instead of X in
(\ref{dKt}), but see the next section for an alternative illuminating
calculation.
Therefore, we just have to verify that
$\rho^{2}V_{\mu}\nabla_{\nu}\Theta^{\mu\nu}{\big|}_{\rho=0}$ is finite, or
equivalently that
$\rho^{3}V_{\mu}\nabla_{\nu}\Theta^{\mu\nu}{\big|}_{\rho=0}=0.$
As
\begin{equation}
V_{\mu}F^{\mu\nu}=\frac{1}{\rho^{2}}(EK^{\alpha}-W^{\alpha}),
\end{equation}
then
\begin{equation}
\label{Vdt}
%% FOLLOWING LINE CANNOT BE BROKEN BEFORE 80 CHAR
4\pi\rho^{5}V_{\mu}\nabla_{\nu}\Theta^{\mu\nu}=-K^{2}{\Big(}2E\rho^{2}\hbox{\Large a}^{2}+3E(1-E^{2})+\rho^{2}(\rho{\dot{\hbox{\Large a}}}.\hbox{\Large a}-E{\dot{\hbox{\Large a}}}_{K}){\Big )},
\end{equation}
and
\begin{equation}
\label{last}
%% FOLLOWING LINE CANNOT BE BROKEN BEFORE 80 CHAR
\lim_{\rho\to0}\rho^{3}V_{\mu}\nabla_{\nu}\Theta^{\mu\nu}=\lim_{\rho\to0}\frac{R^{2}{\Big(}2E\rho^{2}\hbox{\Large a}^{2}+3E(1-E^{2})+\rho^{2}(\rho{\dot{\hbox{\Large a}}}.\hbox{\Large a}-E{\dot{\hbox{\Large a}}}_{K}){\Big)}}{\rho^{4}}.
\end{equation}
Then,  $$A_{0}=R^{2}\Longrightarrow A_{2}=2+{\cal O}(R);\;\;$$
$$B_{0}=2E\rho^{2}\hbox{\Large a}^{2}+3E(1-E^{2})+\rho^{3}{\dot{\hbox{\Large
a}}}.\hbox{\Large a}-\rho E{\dot{\hbox{\Large a}}}_{R}$$
In $B_{0}$ all but $\rho^{3}{\dot{\hbox{\Large a}}}.\hbox{\Large a}$ are terms
of $\tau$-order 2, so
$$B_{0}=2\rho^{2}\hbox{\Large a}^{2}-6\hbox{\Large
a}_{R}-\rho{\dot{\hbox{\Large a}}}_{R}+{\cal O}(R^{3}),$$
$$B_{1}=-5(\rho\hbox{\Large a}^{2}+{\dot{\hbox{\Large a}}}_{R})+{\cal
O}(R^{2}),$$ and $$B_{2}={\cal O}(R).$$ It is not necessary to go further.
Therefore,  according to (\ref{NR}) and (\ref{NP}), we have $$\hbox{p-a=2}$$
and $${\hbox{a=3}},$$ and then, $$p=5>n=4.$$ So, both sides of (\ref{dKt}) are
equally null and there is no contradiction. This is in agreement with the fact
that due to (\ref{BoxA},\ref{J}) and to the antisymmetry of F,
%% FOLLOWING LINE CANNOT BE BROKEN BEFORE 80 CHAR
$$V_{\mu}\nabla_{\nu}\Theta^{\mu\nu}=\frac{1}{4\pi}V_{\mu}F^{\mu}_{\;\;\;\alpha}\nabla_{\nu}F^{\alpha\nu}=V_{\mu}F^{\mu}_{\;\;\alpha}J^{\alpha}=0.$$
\begin{section}
{ Using the divergence theorem}
\end{section}

For the sake of a better understanding of X in eq. (\ref{dKt}) let us work out
its RHS using the divergence theorem. Then we have for the RHS of (\ref{dKt}):
\begin{equation}
\label{X}
\lim_{{\varepsilon_{1}\to0}\atop{\varepsilon_{2}\to\infty}}\int_{\tau_{2}}
%% FOLLOWING LINE CANNOT BE BROKEN BEFORE 80 CHAR
dx^{3}{\Big\lbrace}\Theta^{\mu\nu}\nabla_{\nu}X_{\mu}\;\theta(\rho-\varepsilon_{1})\theta(\varepsilon_{2}-\rho)+X_{\mu}\Theta^{\mu\nu}\nabla_{\nu}\rho[\delta(\rho-\varepsilon_{1})\theta(\varepsilon_{2}-\rho)-\theta(\rho-\varepsilon_{1})\delta(\varepsilon_{2}-\rho)]{\Big\rbrace},
\end{equation}
The explicit dependence on $\nabla_{\nu}X_{\mu}$ makes clear why we cannot just
use K instead of X in (\ref{dKt}): although $\lim_{\rho\to0} K = V$,
$\lim_{\rho\to0} \nabla K \ne \nabla V= -K\hbox{\Large a}.$\\
For working out the first term of (\ref{dKt}) we need:
\begin{equation}
%% FOLLOWING LINE CANNOT BE BROKEN BEFORE 80 CHAR
\nabla_{\mu}K_{\nu}=\nabla_{\mu}(\frac{R_{\nu}}{\rho})=\frac{\eta_{\mu\nu}+K_{\mu}V_{\nu}}{\rho}-\frac{K_{\nu}}{\rho}\nabla_{\mu}\rho,
\end{equation}
\begin{equation}
\label{nablarho2}
W.\nabla\rho\equiv0
\end{equation}
\begin{equation}
K.W=-1
\end{equation}
\begin{equation}
K.\nabla\rho=1+K^{2}E
\end{equation}
\begin{equation}
\Theta^{\mu\nu}\eta_{\mu\nu}=0.
\end{equation}
Then, from (\ref{t'}) and $K^{2}=0$ we have for the upper limit
\begin{equation}
\label{upper}
\lim_{\varepsilon_{2}\to\infty}\int^{\varepsilon_{2}}
%% FOLLOWING LINE CANNOT BE BROKEN BEFORE 80 CHAR
dx^{3}\Theta^{\mu\nu}\nabla_{\nu}K_{\mu}=\lim_{\varepsilon_{2}\to\infty}\int^{\varepsilon_{2}}\frac{d\rho}{\rho^{3}}=-\lim_{\varepsilon_{2}\to\infty}\frac{1}{\varepsilon_{2}^{2}}=0.
\end{equation}
For the lower limit $\nabla_{\nu}X_{\mu}=-K_{\nu}\hbox{\Large a}_{\mu}$ and
then, from (\ref{t'}),
\begin{equation}
4\pi\rho^{4}K.\Theta.\hbox{\Large a}=\hbox{\Large
a}_{K}(K^{2}W^{2}-1)=\hbox{\Large a}_{K}(K^{2}+1)+\rho^{2}\hbox{\Large
a}_{K}K^{2}(\hbox{\Large a}^{2}-\hbox{\Large a}_{K}^{2})-\rho\hbox{\Large
a}_{K}^{2} K^{2}
\end{equation}
So,
\begin{equation}
\label{lower}
\lim_{\rho\to0}4\pi\int\rho^{2}d\rho K.\Theta.\hbox{\Large
a}=-\frac{\hbox{\Large a}_{K}(K^{2}+1)}{\rho} +\hbox{\Large
a}_{K}K^{2}(\hbox{\Large a}^{2}-\hbox{\Large a}_{K}^{2})\rho-\hbox{\Large
a}_{K}^{2}K^{2}ln\rho=0,
\end{equation}
because
\begin{equation}
\lim_{\rho\to0}\frac{\hbox{\Large
a}_{K}(K^{2}+1)}{\rho}=\lim_{\rho\to0}\frac{\hbox{\Large
a}_{R}(R^{2}+\rho^{2})}{\rho^{4}}=0,
\end{equation}
as
\begin{equation}
{\cal O}[\hbox{\Large a}_{R}]+{\cal O}[R^{2}+\rho^{2}]=2+4>4,
\end{equation}
and
\begin{equation}
\lim_{\rho\to0}\hbox{\Large a}_{K}K^{2}(\hbox{\Large a}^{2}-\hbox{\Large
a}_{K}^{2})\rho=0.
\end{equation}
For evaluating the limit of the last term of (\ref{lower}) we consider that
\begin{equation}
{\cal O}[K^{2}\hbox{\Large a}_{K}^{2}]=2={\cal O}[\rho^{2}]
\end{equation}
 to see that
\begin{equation}
\lim_{\rho\to0}\hbox{\Large a}_{K}^{2}K^{2}ln\rho\sim
\lim_{\rho\to0}\rho^{2}ln\rho=0.
\end{equation}
It is important to use the appropriate values of X to have consistent results.
The use, for example, of $X=V$  in the upper limit or of $X=K$ in the lower
limit would produce inconsistent results.

For the second term of (\ref{X}) $X=V$ and then we have from (\ref{t'}) that
\begin{equation}
4\pi\rho^{4}V.\Theta.\nabla\rho=(1+K^{2}E)(W^{2}+E)+\frac{\rho\hbox{\Large
a}_{K}}{2}(1-K^{2}W^{2}).
\end{equation}
Therefore,
\[\lim_{\rho\to0}4\pi\rho^{2}V.\Theta.\nabla\rho   =
\lim_{\rho\to0}{\Big(}\frac{(\rho^{2}+R^{2}+R^{2}\hbox{\Large
a}_{R})(\rho^{2}\hbox{\Large a}^{2}-\hbox{\Large a}_{R}^{2}-\hbox{\Large
a}_{R})}{\rho^{4}}+ \]
\begin{equation}
 \mbox{}+ \frac{{\hbox{\Large a}_{R}}[\rho^{2}+R^{2}-R^{2}(\rho^{2}\hbox{\Large
a}^{2}-\hbox{\Large a}_{R}^{2}+2\hbox{\Large a}_{R})]}{\rho^{4}}{\Big)}=0,
\end{equation}
because
\begin{equation}
{\cal O}[\rho^{2}+R^{2}+R^{2}\hbox{\Large a}_{R}^{2}]+{\cal
O}[\rho^{2}\hbox{\Large a}^{2}-\hbox{\Large a}_{R}^{2}-\hbox{\Large
a}_{R}]=4+2>4
\end{equation}
and
\begin{equation}
{\cal O}[\hbox{\Large a}_{R}]+{\cal O}[(\rho^{2}+R^{2})+\hbox{\Large
a}_{R}^{2}-\hbox{\Large a}_{R}]=2+4>3
\end{equation}
Again we only have consistent results if we use the correct values of X in its
respective limiting situation.

%
%\newpage
\begin{center}
{LIST OF FIGURE CAPTIONS}
\end{center}
\begin{enumerate}
\item Fig. 1.:{The Lienard-Wiechert solutions.}
\item Fig. 2.:{Creation and annihilation of particles in classical physics.}
\item Fig. 3.:{Double limiting process:$\rho\rightarrow0$ along K and
$\tau\rightarrow\tau_{ret}$.}
\item Fig. 4.:{Classical picture of the fundamental quantum process.}
\end{enumerate}
%\end{document}
%\newpage
%
\begin{figure}
%\vglue-30cm
%\epsfig{file=lws.eps,height=10cm,width=10cm}
\epsfxsize=200pt
\epsfbox{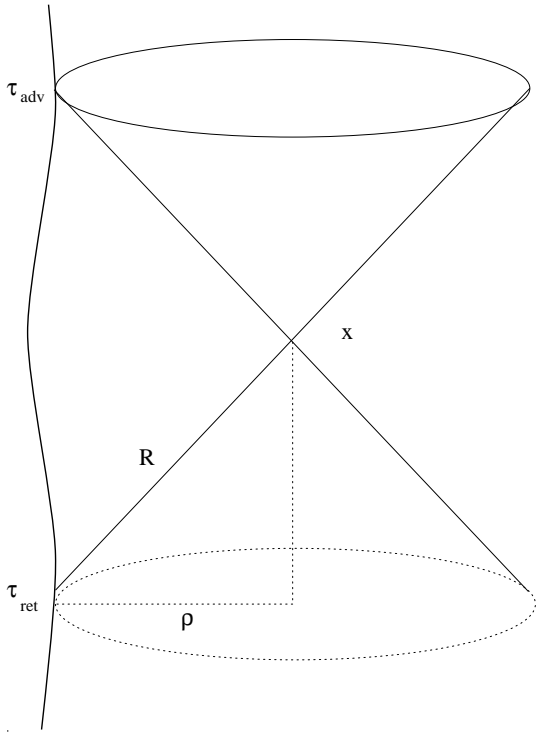}
\caption[Fig. 1.]{The Lienard-Wiechert solutions.}
\end{figure}
%
%\newpage
%
\begin{figure}
\epsfbox{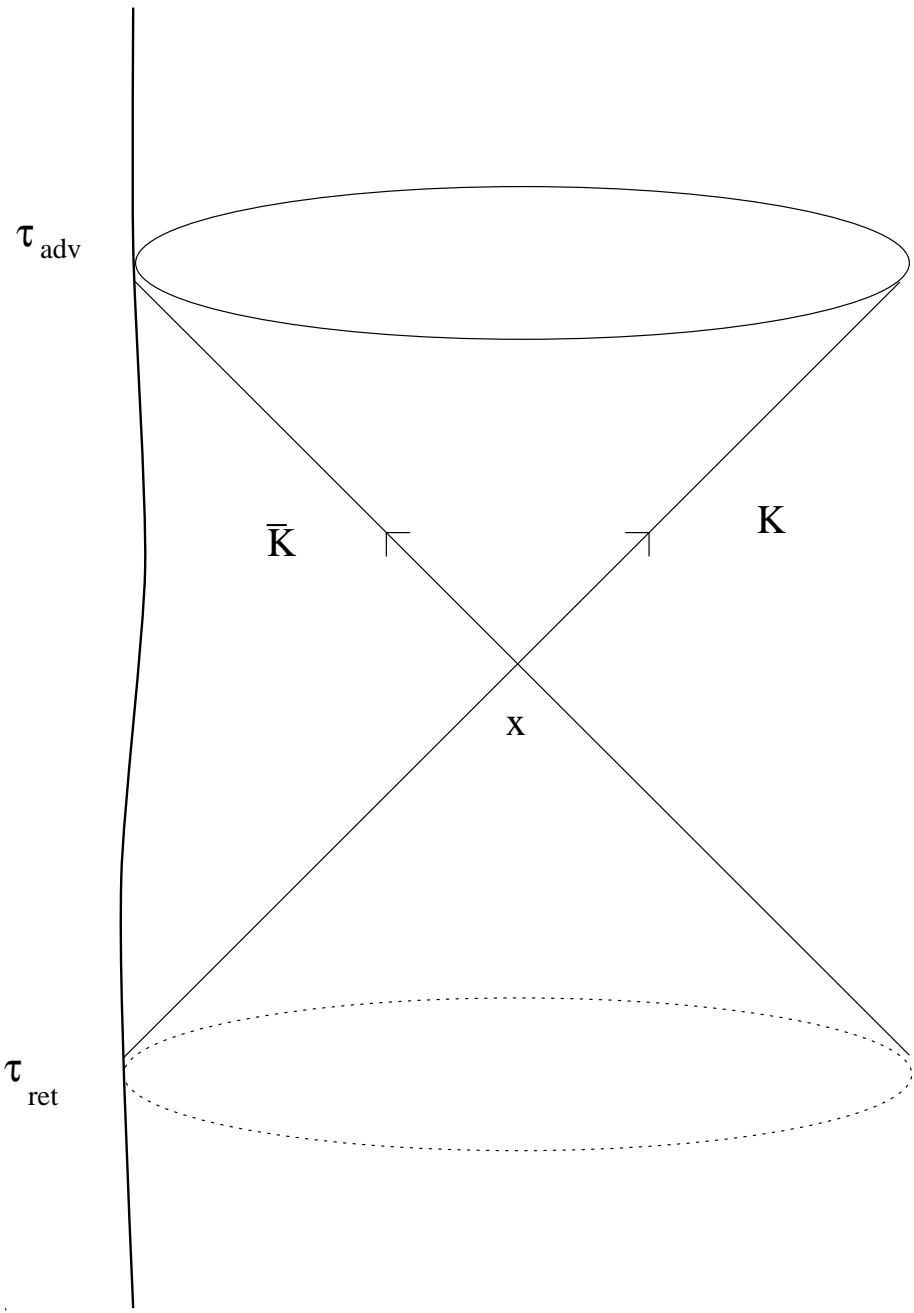}
\epsfxsize=100pt
\caption[Fig. 2.]{Creation and annihilation of particles in classical physics.}
%\vglue-30cm
\end{figure}
%
%\newpage
%
\begin{figure}
\epsfxsize=200pt
\epsfbox{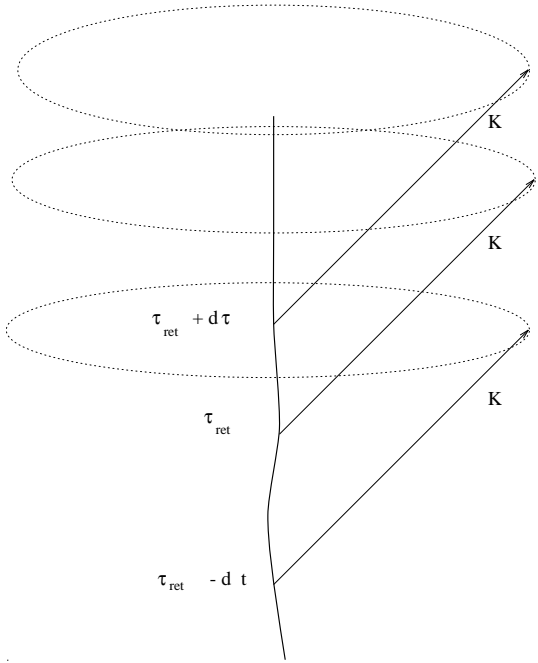}
\caption[Fig. 3.]{Double limiting process:$\rho\rightarrow0$ along K and
$\tau\rightarrow\tau_{ret}$.}
%\vglue-30cm
\end{figure}
%
%\newpage
%
\begin{figure}
\epsfxsize=200pt
\epsfbox{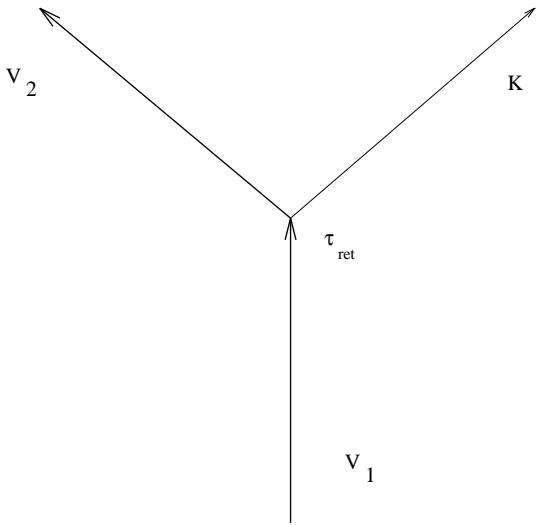}
\caption[Fig. 4.]{Classical picture of the fundamental quantum process.}
%\vglue-30cm
\end{figure}
\end{document}